\documentclass[letterpaper]{article}

\usepackage[T1]{fontenc}

\usepackage{geometry}
\usepackage{setspace}
\usepackage{gensymb}
\usepackage{textcomp} % For \textperthousand and related symbols
\usepackage{url}
\usepackage{achemso}

\usepackage{graphicx}
\usepackage{float}
\newfloat{scheme}{htbp}{los}
\floatname{scheme}{Scheme}
\floatname{chart}{Chart}
\newfloat{graph}{htbp}{loh}

\usepackage{chemformula} % Formulas using \ch{}
\usepackage[version = 4]{mhchem} % Formulas using \ce{}
\usepackage{array}
\usepackage{booktabs}
\usepackage{multirow}
\usepackage{caption}

\usepackage{authblk}
\author[1]{Fernando M. Fernandes*}
\affil[1]{IMCN/NAPS, Universit\'e catholique de Louvain, Louvain-la-Neuve, Belgium}
\author[1]{Beno\^it Hackens}

\title{Model-derived conversion formula for real-time gas monitoring based on chemiresistive sensors}

\date{*Email: fernando.massa@uclouvain.be}

\begin{document}

\maketitle

\begin{abstract}

%  This is an example document for creating \LaTeX{} submissions to the American
%  Chemical Society (ACS) for publication. As ACS does not use \LaTeX{} for
%  typesetting accepted manuscripts, this template does not seek to
%  reproduce the appearance of a published paper.

Chemiresistive gas sensors transduce gas adsorption into changes in the electrical resistance across a pair of electrodes connected by a sensitive layer of material. This type of sensor is used due to its simple operation, high sensitivity, low cost, and convenience for scaled-up manufacturing of microsized devices. The conversion of the electrical resistance to a corresponding gas concentration is often performed through calibration procedures using empirical formulas, overlooking part of the physical phenomena involved in the process, both on the sorption kinetics and on the transduction. Consequently, a direct evaluation of gas concentration is plagued by the response delays and slow recovery intrinsic to these processes. In contrast to this approach, here we first propose a physical model, based on gas-modulated potential barriers, and considering the out-of-equilibrium dynamic response. Based on this model, we derive an original conversion formula able to dynamically convert the resistance changes into a corresponding gas concentration thus eliminating the main drawback related to slow response and recovery. This new strategy is demonstrated for real-time \ch{NO2} gas sensing, using chemiresistors based on oxidized \ch{PbS} nanocrystals. In addition, the broader application of the proposed model and strategy is demonstrated for \ch{NH3} sensing, based on polypyrrole/gold junctions.

\end{abstract}

\section*{Keywords}

gas sensors, chemiresistors, modeling, nanomaterials, environmental monitoring.

%Some journals require keywords: these normally should be given immediately after the abstract.

%\section*{Abbreviations}

%Some journals require a list of abbreviations: these normally should be given immediately after the keyswords (if required).

\section{Introduction}

%This is a paragraph of text to fill the introduction of the demonstration file. 

Today's gas microsensors are widespread in a diverse set of smart applications, including real-time environmental monitoring, indoor air quality control, exhaled breath diagnosis, agriculture, and industrial processes \cite{Zong2025,Mandal2025,Wang2025,Acharyya2024,Song2021}. Among the different gas sensing
principles (e.g., optical, electrochemical, potentiometric, and resistive), gas sensors of the chemiresistive type offers several advantages including high sensitivity, easy operation, ultralow power consumption, affordability, as well as potential for miniaturisation and mass-production (by wafer-level planar integration and batch processing) using modern semiconductor manufacturing technologies \cite{Lakshminarayana2025,Bellitti2022}. Traditionally, semiconductor metal oxides (SMOs) have been widely used, since the introduction of \ch{ZnO}, in the 60’s, and porous \ch{SnO_2} ceramic materials, in the 70’s, as early warning sensors in domestic gas leak detectors \cite{Neri2015}. However, for reactive gases, and strong oxidant as \ch{NO_{2}}, one of the key challenges for a chemiresistive sensor is the recovery owing to strongly adsorbed or absorbed gas molecules. A long-standing issue associated with conventional chemiresistive sensors is the demand for high-temperatures of operation (100-450 \textcelsius) in order to improve performance and accelerate gas desorption \cite{FigaroWeb2025,USTWeb2025,WinsenWeb2025}. Thus, standard chemiresistive sensors must be associated to external heaters, resulting in less compact devices, increased cost, and higher power consumptions (typically in the range of $\sim$100 mW \cite{FigaroWeb2025,USTWeb2025,WinsenWeb2025}) making then unsuitable for mobile and IoT (internet-of-things) applications. Recent approaches to this issue involve the application of micro-heaters \cite{Jung2021,Xu2015}, self-heated (Joule) sensors \cite{Duoc2024,Kim2018}, photoactivated desorption \cite{Lee2025,Li2024}, and great effort has been dedicated to achieve room-temperature operation (i.e. typically involving the design of complex nanomaterials, sensitization with noble metals, e.g. Pt, Au, Pd, and combined strategies) \cite{Bulemo2025,Fan2024,Muthumalai2024,Panigrahi2023,Saruhan2023,Yun2022,Mirzaei2022,Majhi2021,Xuan2020,Jian2020}. However, the adoption of these new strategies for large scale manufacturing has been reluctant, as they convey different degrees of scientific complexities and technical challenges, i.e., trading long-term stability, low-cost and compatibility with modern mass-production manufacturing technologies \cite{Lakshminarayana2025,Lee2023,Xu2015}. A change in the conductance of the sensing layer will manifest itself upon interaction with a reducing gas (e.g., \ch{NH3}, \ch{CO2}, \ch{H2S}, \ch{H2}) or oxidizing gas (e.g., \ch{NO2}, \ch{NO}, \ch{O3}), where the extent of the space-charge layer can be modulated as a result of adsorbed or absorbed molecules acting as electron donors/acceptors \cite{Bulemo2025,Jian2020,Yang2017,Yamazoe2008}. For modeling of the sensor response, the principle of dynamic equilibrium (in the steady state) is usually assumed, i.e. when the adsorption rate is equal to the desorption rate of molecules on the surface of the sensing layer \cite{Hwang1999}. Therefore, one can assume the form of Langmuir \cite{Abbas2015,Joshi2025}, or Wolkenstein \cite{KUMAR2019} equations using empirical (calibrated) constants.  Unfortunately, this approach fails to provide an analytical solution for the dynamic problem, i.e. when the resistance is slowly evolving towards the new equilibrium value after the gas concentration changes. This approach rules out modeling of the sensor dynamic response, in the out-of-equilibrium regime, while long times (several minutes to hours) are effectively necessary for the resistance to stabilize (near the equilibrium) before it can be accurately converted to the corresponding gas-concentration value. In summary, until now, the long reaction/recovery times plaguing chemiresistive sensors operated at room temperature makes then unsuitable for real-time sensing applications, especially for low gas concentrations, in environmental monitoring and indoor air-quality control.

Here we focus on nitrogen dioxide (\ch{NO_2}) gas sensing, i.e. one of the most dangerous air pollutants that can impact human health even at very small concentrations (e.g. the exposure limit in working place, for an 8h-shift, was set to 0.5 ppm (part-per-million) by the EU Commission Directive 2017/164/EU). It is a toxic gas generated from automobiles, combustion sources, and chemical plants, and is a well recognized air pollutant that can threaten public health by triggering asthma and other respiratory diseases \cite{Camilleri2023,Cooper2022}. In recent years, a substantial effort has been dedicated to the development of \ch{NO2} sensor that can operate at ambient temperature \cite{Lee2025,Zhao2025,Fan2024,Muthumalai2024,Xu2024,Brophy2024,Elizabeth2024,Quan2023,Wei2023,Ye2023,Turlybekuly2023,Li2022,Pham2019}. However, despite the efforts on complex material design and merging of different strategies, the use of nanocrystals (NCs), which are easily synthesized in liquid-phase, in combination with ink-printing (or solution-based) deposition techniques, is particularly advantageous for cost effective and scale-up manufacturing using CMOS back-end technology \cite{Lakshminarayana2025,Kwon2024,Barandun2022,Bellitti2022,Sukharevska2021,Liu2021}. In this view, the gas sensing capabilities of lead-sulfide nanocrystals (PbS-NCs) are recognized as it has been demonstrated sensitive towards reducing or oxidizing gases (e.g. \ch{NO_2} \cite{Kwon2024,Li2022,Mitri2020}, \ch{H_{2}S} \cite{Liu2015}, \ch{NH_3} \cite{Liu2016},\ch{CH_4} \cite{Mosahebfard2016}), depending on the specific surface chemistry. A summary of recent results for room temperature operated $\mathrm{NO_{2}}$ sensors based on PbS-NCs is presented in Table \ref{tab.literComp}. The reaction and recovery times of a sensor, $T_{90}$ and $T_{10}$ respectively, are usually defined as the elapsed time between the step-like change in gas concentration and the corresponding response to achieve $90\%$ of the total resistance change towards the new equilibrium value. A general trend, depending on the sorption-kinetics, is that a longer reaction/recovery time is expected when sensing low-concentrations at room-temperature (i.e., especially below 1 ppm). However, in practice, values of $T_{90}$ and $T_{10}$ reported in the literature can be easily underestimated, as the slow sorption kinetics renders impractical to wait several hours for the sensor response to achieve saturation.

In this work, we demonstrate a gas-sensing strategy overcoming the response/recovery time limitations, yielding direct quantitative measurement of gas concentration even during the transient phase before the stabilization of the sensor response. In contrast to the standard approach, this strategy is based on a physical model considering the intergrain potential (Shottky-barrier) configuration, in a sensitive layer made of nanocrystals, and include the out-of-equilibrium dynamic response. Our devices were fabricated using a simple, inexpensive and scalable process, based on the deposition of colloidal PbS-NCs from a liquid-dispersion in water and subsequent heat-treatment at mild-temperatures in a ligand-exchange free process \cite{Hu2023,Vogel2022}. We present and validate a model that can reproduce the dynamic-response when the sensor is exposed to a variable $\mathrm{NO_{2}}$ concentration in air. The device model and the sorption dynamics equations are then combined to derive a conversion formula, that enable a direct conversion from the resistance variation on the devices to a corresponding gas-concentration. The sensing strategy is validated for real-time $\mathrm{NO_{2}}$ gas monitoring, in the range 0.1-0.5 ppm, using a pair of sensors operated at room temperature. In addition, we show that a very similar model also applies to \ch{NH3} sensing with polypyrrole (PPy), another system where the sensing layer is connected to contacts through potential barriers, whose transmission is modulated by gas adsorption. This offers prospects towards a broader applicability of the model and method beyond PbS-NCs. \\

\begin{table}[h!]
\caption{$\mathrm{NO_2}$-sensing performance of PbS-NPs based sensors at room-temperature.} \label{tab.literComp}
\begin{tabular}{lcccc}
\hline
Material  & Conc. (ppm) & Resp. (\%)$^a$ & \ $T_{90}/T_{10} (s)$ $^b$ & Refer. \\
\hline
PbS-QDs treated with $\mathrm{N_{a}NO_{2}}$ & 50  &  92  & 6/28 & \cite{Li2022} \\
PbS-QDs treated with butylamine & 30  &  82  &  12/1560 & \cite{Mitri2020} \\
$\mathrm{MoS_2}$ sensitized with PbS-QDs & 10  &  83.7  & 15/62 & \cite{Liu2020} \\
P3HT dopped with PbS-QDs & 1.0  &  $\sim 50$  & $\sim 600$/$\sim 600$ & \cite{Kwon2024} \\
PbS-QDs capped with iodine & 0.5 &  65.1  & 30/2700 & \cite{Hu2023} \\
PbS-QDs vac-ann. (sample sv) & 0.5 &  58  & 1056/38568 & This work$^c$ \\
PbS-QDs air-ann. (sample sa) & 0.5 &  49  & 6204/54000 & This work$^c$ \\
\hline
\end{tabular}
\text{\textsuperscript{\emph{a}} Resistance drop under exposure to the target gas}. \\
\text{\textsuperscript{\emph{b}} Reaction/Recovery time}. \\
\text{\textsuperscript{\emph{c}} $T_{90}$ and $T_{10}$ estimated using eqs. \ref{eq:transd} and \ref{eq:sorp}.} 
\end{table}

\section{Sensors}

\begin{figure}[h!]
\centering
\begin{tabular*}{\textwidth}{ll}
{$\mathrm{a}$} & {$\mathrm{b}$} \\
{\includegraphics[width=0.5\columnwidth]{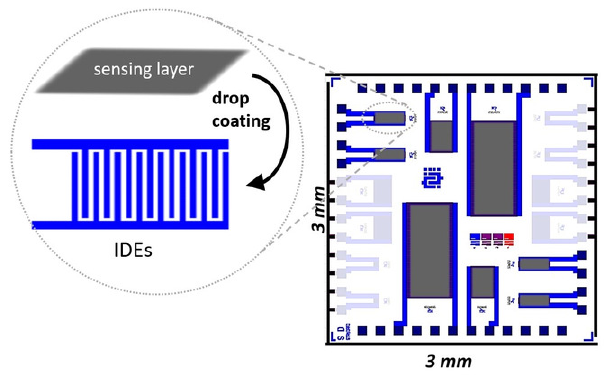}}
 & {\includegraphics[width=0.45\columnwidth]{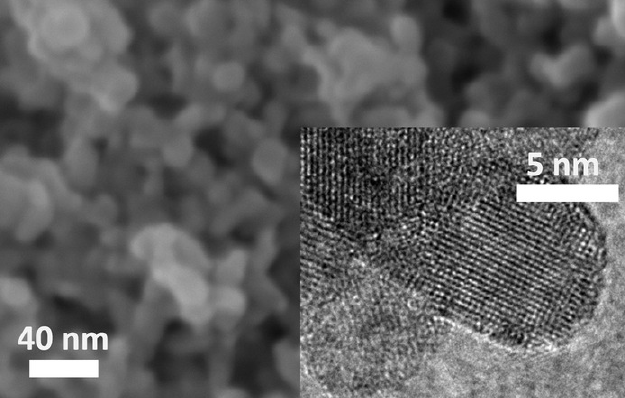}} \\
{$\mathrm{c}$} & {$\mathrm{d}$} \\
{\includegraphics[width=0.5\columnwidth]{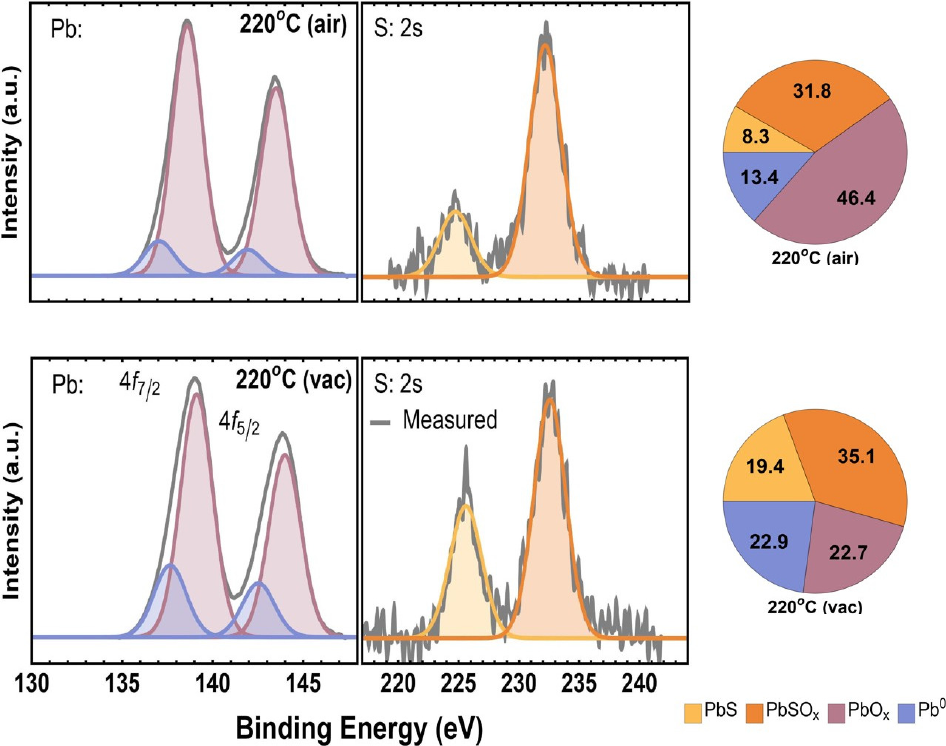}} & {\includegraphics[width=0.45\columnwidth]{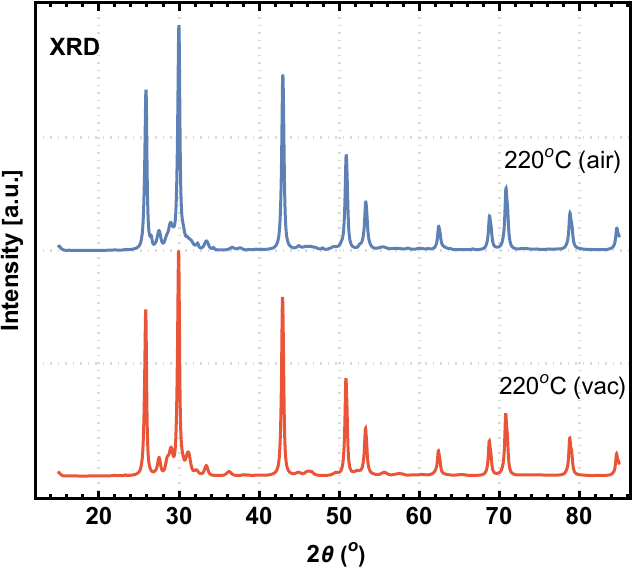}} \\
\end{tabular*}
\caption{(a) A scheme of the sample/die and the geometries of IDEs used for coating with the PbS-NCs. (b) SEM-image showing the morphology of the sensing layer of PbS-NCs. In the inset, is the high-resolution image (HR-TEM) of a single PbS-NC. (c) XRD spectra of PbS-NCs, after thermal-treatment at 220 \textdegree C for 30 minutes in vacuum (in \textit{red}) and after subsequent thermal-treatment at 220 \textdegree C for 30 minutes in ambient-air (in \textit{blue}).}\label{Figure PbS}
\end{figure}

A simple fabrication method was applied based on the drop drying technique, for the deposition of the sensing-layer from a liquid dispersion of PbS-NCs in water directly on top of interdigitated electrodes (IDEs), Fig. \ref{Figure PbS}-a (more details are included in the Experimental section). The PbS-NCs layer has demonstrated tuneable sensitivity when alternating thermal annealing steps after deposition: Heating under vacuum is expected to favor deep interface (donor-like) states related to excess \ch{Pb^0} atoms at the NCs surfaces, while heating in ambient air is expected to promote the formation of a thin polar layer of oxygenated species (e.g. \ch{PbO_{x}} and \ch{PbSO_{x}}) related to surface trap states \cite{Giansante2017,Choi2014}. After the drop drying of PbS-NCs on IDEs-array chips, the samples underwent different thermal treatments in order to tune the surface compositions to promote distinct responses when the samples are exposed to the target-gas. The sample \textsf{\textbf{sv}} was heat-treated in vacuum, at 220 \textdegree C for 30 minutes, while the sample \textsf{\textbf{sa}} was heat-treated in open-air, at 220 \textdegree C for 30 minutes. The annealing is expected to improve the crystallinity of the NCs core and to promote surface modifications, but other effects can also be considered as aggregation and particle fusion. The morphology of the obtained highly porous sensitive layer of PbS-NCs is shown in Fig. \ref{Figure PbS}-b, where in the inset we can verify the presence of the atomic planes at the high-resolution image of a single PbS-NC. 

The surface composition and crystallinity of the PbS-NCs layers, after the heat-treatments (corresponding to sensors \textsf{\textbf{sv}} and \textsf{\textbf{sa}}), were verified respectively by XPS (x-ray photoelectron spectroscopy) and XRD (x-ray diffraction) analysis. For the quantitative XPS-analysis, the peaks were identified in the \textit{lead}-Pb(4f) and \textit{sulfur}-S(2s) regions of the narrow spectra. The surface amounts of lead-sulfide (\ch{PbS}), sulfates (\ch{PbSO_{x}}), oxides (\ch{PbO_{x}}), and neutral-lead (\ch{Pb^{0}}), were quantified and the results are presented in Fig. \ref{Figure PbS}-c. As expected, heating the deposited PbS-NPs layer in open-air, as for sample \textsf{\textbf{sa}}, favors the formation of higher amounts of oxygenated species, 78.2 \%, against 57.8 \% for the layer heated in vacuum. Another fundamental difference to highlight, is the presence of sulfur-richer surface on the sample treated in vacuum, i.e. 54.5 \% of sulfur-containing species compared to 40.1 \% on the sample heated in open-air. The effect of the heat-treatment on the crystallinity of PbS-NCs was confirmed by XRD-analysis and the results are shown in Fig. \ref{Figure PbS}-d. A sample heated in vacuum at $\mathrm{220^{o}C}$ for $\mathrm{30\ minutes}$ presented a mean crystallite size of $\mathrm{34 \pm 3 \ nm}$ (as determined from the XRD-spectrum). After a subsequent step of heating at $\mathrm{220^{o}C}$ for $\mathrm{30\ minutes}$ in ambient-air, one observes a minor decrease of the mean crystallite size to $\mathrm{31 \pm 3 \ nm}$, probably due to surface oxidation with the formation of oxygenated species \cite{MOZAFARI2012,Sadovnikov2011}. Overall, both sensors presents similar surface morphology and crystallite sizes after the respective post-deposition processing.

\section{Response}

\begin{figure}[h!]
\centering
\begin{tabular*}{\textwidth}{ll}
{$\mathrm{a}$} & {$\mathrm{c}$} \\
\includegraphics[width=0.5\columnwidth]{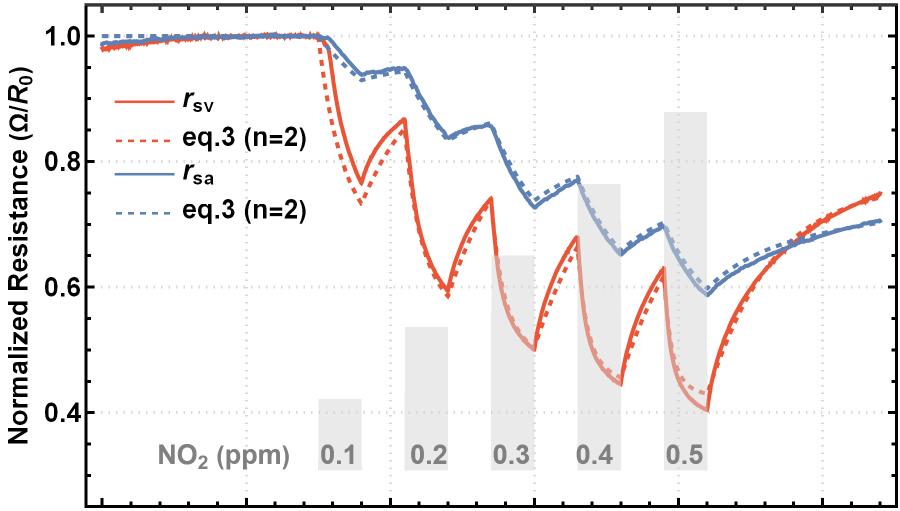} & \includegraphics[width=0.5\columnwidth]{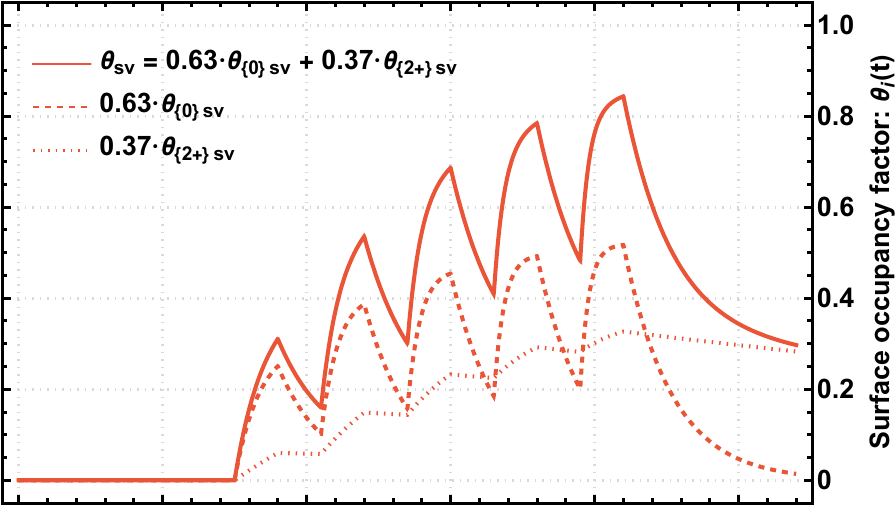} \\
{$\mathrm{b}$} & {$\mathrm{d}$} \\
\includegraphics[width=0.5\columnwidth]{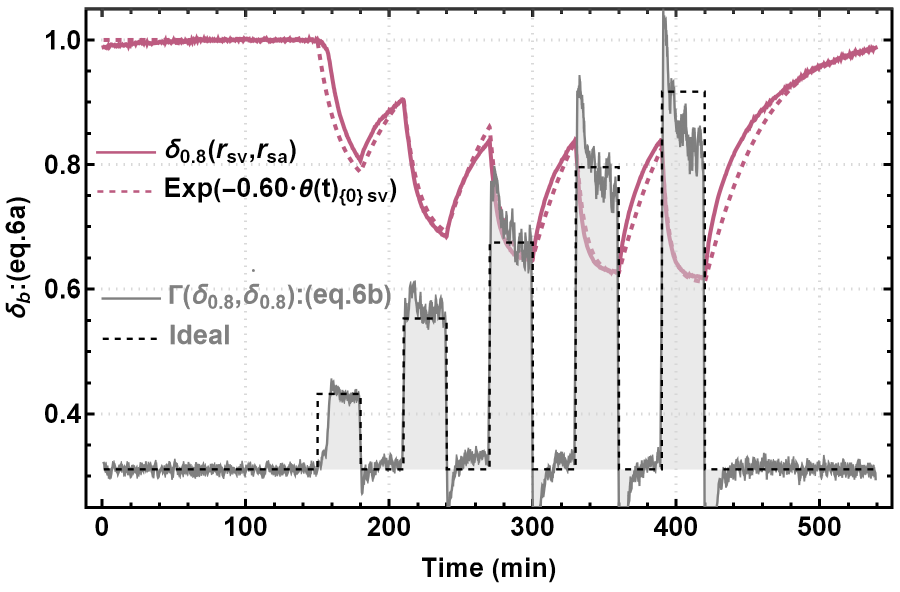} & \includegraphics[width=0.5\columnwidth]{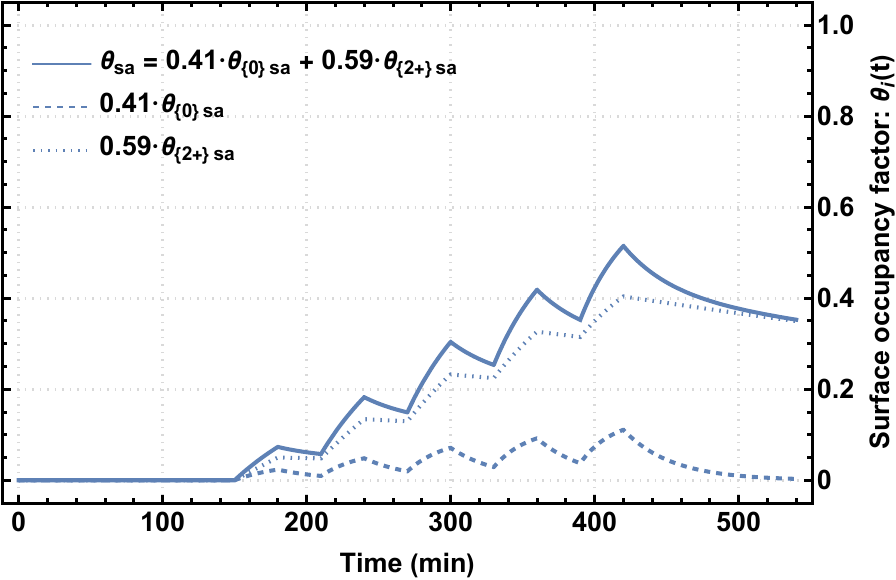} \\
\end{tabular*}
\caption{(a) Normalized resistance variation of sensors \textsf{\textbf{sv}} and \textsf{\textbf{sa}} when different concentrations of \ch{NO_2}-gas are mixed (in intervals of 30-minutes) to a constant flow (2-liters/minute) of synthetic-air with constant relative humidity (44\%). The \textit{red lines} corresponding to sample \textsf{\textbf{sv}} and the \textit{blue lines} corresponding to sample \textsf{\textbf{sa}}. The \textit{continuous} lines corresponds to the experimental data, and the \textit{dashed} lines are fittings to the data using the model, with the parameters in Table \ref{tab.ModParam} substituted in (\ref{eq:transd}) and (\ref{eq:sorp}). (b) Validation of the DS-strategy: In \textit{purple colour}, the comparison between the continuous and the dashed lines allows to confirm the validity of relation (\ref{eq:dconva}) when applied to the resistance data shown in panel (a). In panels (c) and (d), the corresponding surface occupancy functions of each sensor during the test, $\theta_{sv}(t)$ and $\theta_{sa}(t)$, after solving (\ref{eq:sorp}).}
\label{Fig.resp}
\end{figure}

For the \ch{NO_{2}}-gas sensing test, a sequence of 5-pulses were applied with step-like \ch{NO_2} gas concentrations ranging from 0.1-to-0.5 ppm, in a constant flow of synthetic-air, and the electrical resistances of samples \textsf{\textbf{sv}} and \textsf{\textbf{sa}} were recorded over-time (with sensors operated at room temperature). The temporal evolution of the respective normalized resistances, $r_{sv}$ and $r_{sa}$, are reported in Fig.\ref{Fig.resp}-a. The distinct curves observed for each sample is attributed to the difference on the surface composition (Fig.\ref{Figure PbS}-c), corresponding to the distinct thermal treatment applied after deposition (in vacuum and ambient-air, respectively). Both samples responded significantly to the presence of \ch{NO_2} gas for all tested concentrations. The slower recovery observed on sample \textsf{\textbf{sa}}, compared to sample \textsf{\textbf{sv}} (Tab.\ref{tab.literComp}), can be generally associated to a dominant presence of interaction sites with higher binding energy at the surface. Further, the slower reaction observed on sample \textsf{\textbf{sa}} (\textit{blue curve}, in Fig. \ref{Fig.resp}-a), is the manifestation of a surface with lower sticking probability in comparison to sample \textsf{\textbf{sv}}. This can be related to longer surface diffusion lengths, as in a partially passivated surface.

\subsection{Mechanism}

In a sensing layer made of PbS-NCs, the PbS cores are encapsulated by an oxidized surface layer
that is reactive to \ch{NO_{2}} \cite{Fernandes2026}. The sensing layer is formed by an ensemble of connected NCs in which conductivity is assumed to be dominated by the built-in intergranular potential barriers, formed between adjacent nanocrystals, as depicted in Fig.\ref{Fig.Scheme-device}-a, and sometimes referred as double-Schottky barriers \cite{Mirzaei2022,Jian2020}. The proposed sensing mechanism involves the tuning of the space-charge region (electron-depletion/hole-accumulation in n/p-type material) close to the surface of the nanocrystals (represented in white around the NCs, in Fig.\ref{Fig.Scheme-device}), when the sensing layer interacts with molecules of the target gas in the atmosphere leading to a modulation of the interface potentials.

Regarding the \ch{NO_{2}} sensing mechanisms on oxidized PbS-NCs, we consider the reactivity and sorption dynamics defined by the surface composition, more precisely the concentration ratio of lead in the oxidation states \ch{Pb^{0}}, \ch{Pb^{2+}}, and \ch{Pb^{4+}}, as demonstrated in the companion paper \cite{Fernandes2026}. Overall, the sensing mechanism can be associated to the presence of surface amounts of \ch{Pb^{0}} and \ch{Pb^{2+}} that are associated to the formation of adsorbed nitrate via surface-mediated reactions involving \ch{PbO} and \ch{Pb-OH} (formed by hydroxylation of metallic-lead) available on the surface \cite{Fernandes2026}. Thus, one can consider the presence of two-types of interaction sites corresponding to the surface amounts of \ch{Pb^{0}} and \ch{Pb^{2+}}, while the presence of surface amounts of \ch{Pb^{4+}} are associated to passivated site\cite{Fernandes2026}. However, here, we should emphasize that a more detailed knowledge and control of the precise adsorption/reaction mechanism on the surface was not necessary to develop the method described hereafter. Nevertheless, a better understanding of the atomic-level mechanism would be desired to optimize the method. To this end, we refer to the related work co-authored by both authors of the present work, where a much more in-depth investigation of the mechanisms is proposed, involving density-functional-theory-based simulation results (Fernandes et al., 2026)\cite{Fernandes2026}.

\section{Model}

\begin{figure}[h!]
\centering
\begin{tabular*}{\textwidth}{ll}
{$a$} & {$b$} \\
{\includegraphics[width=0.5\textwidth]{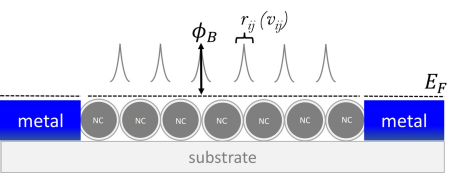}}  & {\includegraphics[width=0.5\textwidth]{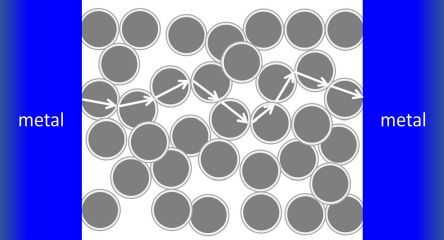}} \\
\end{tabular*}
\caption{Configuration of a sensing device formed by nanocrystals. (a) Side view representation of intergranular potential barriers ($\phi_{B}$) that are formed between adjacent nanocrystals (NCs). (b) Top view configuration of a device based on a random network of nanocrystals connecting the two electrodes, where the conductive path is formed by random percolation chains (\textit{white arrows representing the charge path)}. The electron-depletion layer on the surface of n-type particles (or a hole-accumulation layer for p-type particles) is represented in white around the NCs. }\label{Fig.Scheme-device}
\end{figure}

In the case of a sensing layer formed by semiconductor NCs, a simple approach is to assume each NCs as conducting cores (particles) embedded in an insulating medium. The NCs assembly can be modeled as a series-parallel random combination of connected NCs forming a pathway of resistances bridging the pair of electrodes \cite{Zabet-Khosousi2008}. The configuration is represented in Fig.\ref{Fig.Scheme-device} for a typical sensing device formed of nanocrystals, with ohmic contacts on both sides. For a random network of nanocrystals, i.e. formed by spin coating or ink-based deposition technique, the conductive path is formed by random percolation chains bridging the pair of electrodes, as represented in Fig. \ref{Fig.Scheme-device}-b (\textit{white arrows}). We assume each pair of connected NCs (\textit{i, j}) in the percolation network contributing with a junction resistance $r_{ij}$, in the form of eq. (\ref{eq:aprij}) below, where $\phi_{B_{ij}}$ is the height of the inter-grain potential barrier, $S$ is the junction area, $A^{*}$ is the Richardson constant, $T$ is the temperature and k is the Boltzmann constant. The total resistance of the sensing layer, $R$, depends on the resulting configuration of the disordered-network formed by the randomly cross-linked pathways of connected NCs. For a sufficiently narrow distribution of $r_{ij}$, we assume that the total resistance is then proportional to the junction resistance (\ref{eq:R}). The relations in (\ref{eq:Rrij}) were validated by electrical resistance versus temperature measurements, and the results are included in the supporting information.

\begin{subequations}
\begin{align}
r_{ij} = \frac{2k}{S_{ij}A^{*}T}\cdot \exp{\left(\frac{\phi_{B_{ij}}}{kT}\right)}\label{eq:aprij}  \\
R \propto r_{ij} \label{eq:R}
\end{align}
\label{eq:Rrij}
\end{subequations}

The sensor response is modeled by defining a \textit{transfer function}, $f_{tr}$, that allows to express its electrical resistance, $R(t)$, in response to an atmosphere containing a variable concentration, $\Gamma(t)$, of the target gas: $R(t) = f_{tr}(\Gamma(t)) $. It depends on how the adsorption, or reaction, dynamics of molecules on the surface of the sensitive material (the sorption dynamics) will affect its electrical resistance. The chemiresistive behavior is generally observed as the change in the electrical resistance of a sensing-layer as a function of the surface occupancy factor, $\theta$, that corresponds to the fraction of occupied interaction-sites $(0<\theta<1)$. The way the resistance is affected by $\theta$ is described by the \textit{transduction function}, $f_{td}$: $R(\theta) = f_{td} \left( \theta \right) $. On the other side, the sorption dynamics is described by $\theta(t)$, in terms of the time-varying gas-concentration, $\Gamma(t)$, and is defined by the so called \textit{receptor function}, $f_{re}$: $\theta(t) = f_{re}(\Gamma(t))$. Therefore, the \textit{transfer function}, $f_{tr}$, is a composite function ($f_{tr} \rightarrow f_{td}  \circ f_{re} $) and the electrical resistance is defined as $R(t) = f_{td}(f_{re}(\Gamma(t))) $. Next, for the barrier height as a function of $\theta$, we assume equal contributions from the adsorbed molecules leading to a linear relation: $\phi_{B} = (E_{g}-\phi_{0})-\alpha\theta$, where $E_{g}$ is the semiconductor gap-energy, $\phi_{0}$ is the energy level of the interface trap states at neutrality, and $\alpha$ is the contribution to the barrier height when all surface interaction sites are occupied (i.e., for $\theta = 1$). As a generalization, in the case of existing \textit{\textquotesingle n\textquotesingle}-different types of interaction-sites on the surface, and considering the adsorbed phase behave ideally (i.e., there is no intermolecular interactions on the surface), we assume:

\begin{equation}\label{eq:potbar}
   \phi_{B}=\left(E_{g}-\phi_{0}\right)-{\sum_{i}^{n} \alpha_{i}\theta_{i}}
\end{equation}

When the sensor operates at constant temperature and humidity conditions, we can express the \textit{transduction function} in a convenient form, in terms of a baseline resistance, $R_{0}$ (i.e. the condition where the sensor is exposed to a neutral atmosphere), and write:

\begin{equation}\label{eq:transd}
   R = R_{0} \cdot \exp{\left(- \frac{ {\sum_{i}^{n} \alpha_{i}\theta_{i}}}{kT} \right)}
\end{equation}

Finally, to model the response of our sensors, we assume (empirically) that the surface interactions with the gas-molecules are governed by a simple first-order sorption kinetics described by (\ref{eq:sorp}) \cite{Gardner1995}, where \textit{\textquotesingle i\textquotesingle} corresponds to the type of interaction, i.e. $i=\{0\}$ or $i=\{2+\}$, respectively to the presence of metallic-lead (\ch{Pb^{0}}) and lead oxide (\ch{Pb^{2+}}). The first term ($A_{i}$) is related to adsorption and the second term ($D_{i}$) is related to desorption process. In principle, the parameters $A_{i}$ and $D_{i}$ can be obtained experimentally, by fitting the sensor response using (\ref{eq:transd}), provided that (\ref{eq:sorp}) can be solved to find $\theta_{i}(t)$ as a function of the experimental gas concentration $\Gamma(t)$.

\begin{equation}\label{eq:sorp}
\frac{\partial\theta_{i}}{\partial t} = A_{i} \Gamma (1-\theta_{i}) - D_{i} \theta_{i}
\end{equation}

Thus, the model response curves (dashed lines), in Fig.\ref{Fig.resp}-a, correspond to fitted curves using eqs. (\ref{eq:transd}) and (\ref{eq:sorp}). The profile of the predefined \ch{NO_{2}} gas concentration used during the gas sensing test (shadowed areas, in Fig.\ref{Fig.resp}-a) was substituted in eq. (\ref{eq:sorp}), for both surface interaction types, and solved simultaneously with respect to (\ref{eq:transd}). The surface occupancy functions, $\theta_{\{0\}}(t)$ and $\theta_{\{2+\}}(t)$, were then found (numerically), together with their associated parameters $\alpha_{\{0\}}$ and $\alpha_{\{2+\}}$, for each sensor (\textsf{\textbf{sv}} and \textsf{\textbf{sa}}), and the resulted occupancy functions are shown in Figs.\ref{Fig.resp}-c,d. Finally, a remarkable correspondence is verified between the experimental response (solid lines) and the model (dashed lines), in Fig.\ref{Fig.resp}-a, using the parameters presented in Table \ref{tab.ModParam}.

\begin{table}[h!]
\centering
\caption{Model parameters for fitting of (\ref{eq:transd}) and (\ref{eq:sorp}) in Fig.\ref{Fig.resp}-a} \label{tab.ModParam}
\begin{tabular}{lcccc}
\hline
\textbf{$\mathrm{\mathbf{sv}}$} & $A_{\mathrm{i}}$ & $D_{\mathrm{i}}$ & \ \ $\alpha_{\mathrm{i}}/kT$ & $R_{0}=7.1 M \Omega $ \\
\hline
$ \theta_{\{0\}}$  &  0.28  &  0.030 & 0.63 \\
$ \theta_{\{2+\}}$  &  0.06  &  0.0012 & 0.37 \\
\hline
\textbf{$\mathrm{\mathbf{sa}}$} & \ & \ & \ \ \ & $R_{0}=2.7 M \Omega $ \\
\hline
$ \theta_{\{0\}}$  &  0.03  &  0.030 & 0.41 \\
$ \theta_{\{2+\}}$  &  0.03  &  0.0012 & 0.59 \\
\hline
\end{tabular}
\end{table}

\section{Conversion formula}

In the case of a single type interaction site on the surface, we can find an analytical formula to directly convert, in real-time, the sensor response, $R(t')$, to a gas concentration, $\Gamma(t')$, i.e., by taking the time-derivative of (\ref{eq:transd}), substituting ${\partial\theta}/{\partial t}$ by (\ref{eq:sorp}) and using $\theta = \left(kT / \alpha \right) \ln{(R_{0}/R)}$ we find the conversion formula in \ref{eq:sconv}. The set-of-parameters, $A$, $D$, $\alpha$ and $R_{0}$, can be obtained from calibration by fitting the conversion formula (\ref{eq:sconv}), using the experimental curve, $\Gamma(t)$, and the response curve, $r(t)$, of the sensor:

\begin{subequations}
\begin{align}
r=R/R_{0} = r(t) \label{eq:sconva}  \\
\Gamma (r,\dot r) =-\frac{D \ln r+\frac{\dot r}{r}}{A\left[\frac{\alpha}{kT}+\ln r\right]} = \Gamma (t) \label{eq:sconvb}
\end{align}
\label{eq:sconv}
\end{subequations}

However, the conversion formula (\ref{eq:sconv}) can not be applied in the presence of distinct types of interaction sites, as in our sensor-pair based on PbS-NCs. Therefore, a different sensing strategy using samples \textsf{\textbf{sv}} and \textsf{\textbf{sa}} was developed. In the following, we propose a suitable sensing scheme based on the relative resistance measurement, using a pair of devices with distinct combinations of different types of interaction sites on the surface. In the case of n-different interaction sites we define the relative resistance function $\delta_{b}(r_{sv},r_{sa})$, for the double-sensor strategy (DS-strategy), where $r_{sv}$ and $r_{sa}$ are respectively the normalized resistances of the sensors in the pair, and one must find a pair of constant-factors, $a$ and $b$, to satisfy the relation in (\ref{eq:dconva}) for some surface occupancy function $\theta_{\{i\}sv}$. If (\ref{eq:dconva}) is found valid, the gas concentration can be determined in real-time using the DS-strategy, i.e., taking the conversion formula given in (\ref{eq:sconv}) and substituting $r \Leftrightarrow \delta_{b}$, and $\left(\alpha / kT, A, D \right) \Leftrightarrow (a, A_{\{i\}sv}, D_{\{i\}sv})$ to obtain (\ref{eq:dconvb}).

\begin{subequations}
\begin{align}
\delta_{b}(r_{sv},r_{sa}) = \frac{r_{sv}}{r_{sa}^b}  \approx  \mathrm{exp}{\left( -a \cdot \theta_{\{i\}sv} \right)}\label{eq:dconva}  \\
\Gamma(\delta_{b}, \dot \delta_{b}) =-\frac{D_{\{i\}sv} \ln \delta_{b}+\frac{\dot \delta_{b}}{\delta_{b}}}{A_{\{i\}sv}\left[a+\ln \delta_{b}\right]}\label{eq:dconvb}
\end{align}
\label{eq:dconv}
\end{subequations}

Therefore, we found (\ref{eq:dconva}) satisfied for $\theta_{\{0\}sv}$ when $a=0.60$ and $b=0.80$. The good correspondence between the experimental curve $\delta_{0.8}(t)$ and the function $\exp{\left( {-0.60 \cdot \theta_{\mathrm{\{0\}sv}}(t) } \right)}$, as established in (\ref{eq:dconva}), can be verified in Fig.\ref{Fig.resp}-b (in purple). In Fig.\ref{Fig.resp}-b, the gray curve (shaded area) shows the converted \ch{NO_2} gas concentration using (\ref{eq:dconv}), with the values of $A_{\{0\}sv}$, $D_{\{0\}sv}$, $R_{0(sv)}$ and $R_{0(sa)}$ extracted from the sensor model (Table \ref{tab.ModParam}). The sequence of measured data-pairs, $r_{sv}$ and $r_{sa}$, with 1-minute interval between consecutive pairs, was used to calculate the values in $\delta_{0.8}(r_{sv},r_{sa})$ (\ref{eq:dconva}). During continuous measurement, the last and third-to-last data-pairs are used to approximate the time-derivative $\dot{\delta}_{b}$, in (\ref{eq:dconvb}), whereas the mid-point (second-to-last) corresponds to the time-label, $t$, defined as the instant for which the gas concentration is reported (in, Fig.\ref{Fig.resp}-b). Using this measurement-protocol, in principle, one can obtain the gas-concentration at the instant $'t'$ with a time delay similar to the approximate sampling-rate (i.e. 1-minute in our case). It is possible to verify the remarkable agreement using the DS-strategy, between the converted gas concentration, $\Gamma(\delta_{0.8},\dot \delta_{0.8})$ (the gray-shaded area), and the ideal gas concentration in the test chamber (black-dashed line), in Fig.\ref{Fig.resp}-b. Despite the long reaction and recovery times characterizing the individual sensors in the pair (Tab.\ref{tab.literComp}), the $\mathrm{NO_{2}}$ concentration is reported without delay for every step-like gas concentration added in the chamber during the test (the small delay observed in the first pulse can be attributed to a experimental factor, related to the low gas flow used in the first pulse). The reported values at lower concentrations (0.1 and 0.2 ppm) seem more precise and accurate while at higher concentrations the reported values are susceptible to the presence of measurement artifacts when the gas concentration changes abruptly (this effect can be a manifestation of non-optimized parameters, or the approximation related to the simple sorption dynamics assumed in eq. \ref{eq:sorp}). This results allows us to validate the conversion formula directly from the model obtained for the individual sensors in the pair. In addition, the sensing performance using the same conversion formula, i.e. with the same set-of-parameters used in Fig.\ref{Fig.resp}-d, was further validated for \ch{NO_{2}} gas concentrations a for completely independent resistance-variation data-sets (the results are shown in the supporting information).

To confirm the applicability of our model beyond PbS NCs networks, we present data for polypyrrole sensors, based on the conversion formula in (\ref{eq:sconv}), derived for single-type interaction site. This approach was experimentally validated for ammonia (\ch{NH3}) sensing by polypyrrole (PPy) sensors (see the supporting information for details of sample fabrication and experimental data). In this case the sensing principle involves the modulation of the Schottky barriers that are formed at the junction between a film of highly (p-type) doped polymer and a metal with large work function, i.e. PPy (electrochemically polymerized) deposited on gold-electrodes \cite{NGUYEN1999}.

\section*{Conclusion}

The physical model of chemiresistive sensors laid out in this work significantly improves and fundamentally changes the gas sensing strategy based on this type of sensor. The model, based on adsorption-modulated intergrain potentials, correctly describes the dynamic response of gas sensors based on PbS-NCs when exposed to concentrations of \ch{NO_2} gas in air, in the out-of-equilibrium condition. From this model, a conversion formula was derived to enable the quantitative sensing of a gas-concentration in real-time from the relative differential-resistance variation in a sensor-pair. We validate a sensing scheme to enable real-time \ch{NO_2} gas monitoring using a simple pair of microsensors operated at room temperature. Moreover, this new sensing strategy was validated for monitoring of low concentrations of \ch{NO_2} in air, in a range where it starts to threaten human health and the environment. Based on this approach, optimized calibration and measurement protocols can be envisaged to enable real-time gas monitoring with even higher accuracy. In addition, the device fabrication process is simple, cost-effective, and suitable for large-scale production of microsized gas-sensors.

The application of this new sensing strategy allows to compensate the sorption dynamics on the sensing layers and yield the instant value of gas concentration in real-time (i.e. in a way that is independent of the response/recovery time characteristic of individual sensors). This sensing scheme eliminates one of the main drawbacks of chemiresistive devices when operated at room temperature and unlocks their application in smart and mobile sensors for precision gas detection and monitoring in real-time using microsized devices. In addition, the sensing strategy described here is not limited to PbS-NCs, or sensing of \ch{NO_2}. In the supporting information, this strategy is validated for application on \ch{NH_3} sensing, based on a single sensor made of conductive polymer (polypyrrole), demonstrating the broader applicability of our model to other systems involving potential barriers whose transmission is modulated by gas adsorption or reaction.

\subsection*{METHODS}\label{methods}

\textbf{PbS nanoparticles:} The PbS-NCs were obtained via a simple wet chemical route in water, at room-temperature, using \ch{Pb(NO3)2} (Merk product number: 203580-10G) and \ch{Na2S} (Merk product number: 407410-10G) as precursors and 2-Mercaptoethanol (Merk product number: M6250-100ML) as the surfactant \cite{Mosahebfard2016,Nabiyouni2012}. Initially, 50 mL solution of \ch{Pb(NO3)2} with concentration 0.1M is poured into a triple-neck round bottom flask, and a volume of 100 mL of 2-Mercaptoethanol solution, with concentration 0.1M, is then added dropwise, followed by 50 mL of \ch{Na2S} solution, with concentration of 0.1M, that was slowly added. The mixture was vigorously stirred until the color of the final suspension turned black. The process was done at room-temperature using Milli-Q water in the solutions (that was previously degassed in argon for 15 minutes). A constant pressure of argon was permanently maintained, in the flask, in all stages during the synthesis. As a result, this synthesis protocol produces a dispersion of PbS-NCs with concentration of approximately 5 mg/mL. \\ \\
\textbf{Interdigitated electrodes:} Commercial IDEs-array chips (CMOSEnvi\textsuperscript{\textregistered}) were used as substrates for samples production, supplied by the company VOCSens \cite{VOCSens}. A total of $8$x IDEs on each chip were coated with PbS-NCs to form individual sensors. The IDEs are formed by thin-film strips consisting of Ti:Au layers (200:2000 \text{\AA} of thickness), with 2 $\mathrm{\mu m}$ wide and spaced by a 5 $\mathrm{\mu m}$ gap. The metal-strips are deposited on top of a $\mathrm{SiO_2}$ insulating layer. The area of each micro-electrode pair ranged from 50 $\mathrm{\mu m}$ x 350 $\mathrm{\mu m}$, on the smallest, to 500 $\mathrm{\mu m}$ x 1050 $\mathrm{\mu m}$, on the biggest pair (Fig.\ref{Figure PbS}a). All the samples used in this work were fabricated using the IDE with an area of  $50\mu m$ x $350\mu m$, corresponding to the smallest IDE (as detailed in the inset of Fig.\ref{Figure PbS}a). \\ \\
\begingroup
\textbf{Deposition of PbS-NCs on the IDEs:} For the sensors fabrication, the deposition of PbS-NCs was based on the drop drying method. For the deposition, the liquid-dispersion (Ink) of PbS-NCs in water (5mg/mL, as obtained after the synthesis) was subjected to an ultrasonic bath at mild-power (70-80W) before being filtered in a sequence using syringe-filters with 5.0 µm, 2.7 µm and 1.2 µm pores. The filtered ink was then loaded in a micro-fine syringe, with a 0.3 mL reservoir. The syringe needles must be cut vertically to form a perfect perpendicular cut at the extremity of the needle. The diameter of the needle is approximately ~400 µm. With this technique a droplet with sub-nanoliter volume can be formed (manually) in a semi-controlled way. The syringe is mounted on an old-wire bonder station to provide stability and mechanical precision on the 3-axis. In addition, the Wire bonder is equipped with a stereo microscope for sample visualization. The sample is placed on a sample holder, that is equipped with a ceramic heater and thermocouple. During the drop casting the sample is heated at 65-80 \textcelsius \ to increase the water-evaporation rate and promote the formation of a homogeneous coating layer. After the drop casting, the samples are vacuum-dried at 70 \textcelsius \ for 2 h. \\ \\
\endgroup
\begingroup
\textbf{XPS:} The XPS analysis was performed on deposited layers of PbS-NCs on Au-coated \ch{SiO2} substrates. The XPS measurements were performed with a PHI 5000 VersaProbe III Photoelectron Spectrometer (Physical Electronics (USA)), equipped with a monochromatized micro focused Al K$\alpha$ X-ray source, powered at 50 W. The pressure in the analysis chamber was kept around 10-6 Pa. The angle between the surface normal and the axis of the analyzer lens was 45°. High resolution scans of the Pb4f and S2s photoelectron peaks were recorded from a spot diameter of 200 $\mu$m using pass energy of 13 eV and step size of 0.1 eV. Charge stabilization was achieved thanks to a combination of Argon and electron guns. Data treatment was achieved using the CasaXPS software (Casa Software Ltd, UK). The Shirley algorithm was applied to remove the background and the shape of the spectral peaks were modeled (fitted) using Gaussian-Lorentzian product formula (with 85\% Gaussian and 15\% Lorentzian). \\ \\
\endgroup
\begingroup
\textbf{XRD:} The ink of PbS-NCs was deposited by the drop-drying method on zero-background Si Holders. The XRD spectrometer (Bruker D8 Discover) was operated at 30 mA, using Cu K$\alpha$ (Ni-filtered) radiation source with a wavelength of 1.5406 Å, and equipped with a LynxEye detector. The goniometer was equipped with a fixed divergent slit opening of 0.6 mm and Soller slit with opening angle of 2.498° on the primary-beam, and anti-scatter slit opening of 6.76 mm and Soller slit opening of 4° on the secondary-beam. Data was acquired between 35° and 85° and collected with a step interval equal to 0.00831° over 2$\theta$, with a scan speed of 0.1 sec/step. The evolution of the crystallinity was determined after each step of thermal treatment by analyzing the XRD spectra, applying the Rietveld refinement technique using Profex software, to characterize and quantify the composition of crystalline phases (Fig.\ref{Figure PbS}c-d). A total of 4-steps of thermal processes were applied sequentially to the XRD samples: \textbf{(1)} The first step of the thermal process consisted in a vacuum-assisted annealing at 150 $\celsius$ for 40 minutes. \textbf{(2)} The second step consisted of increasing the temperature to 180 $\celsius$, in vacuum, and waiting for 30 minutes before the next step. \textbf{(3)} In the third step, the temperature was increased again, in vacuum, up to 220 $\celsius$ and maintained for 30 minutes. Finally, in the last step \textbf{(4)}, the sample was removed from the vacuum-assisted annealing system and placed on a hot-plate, in open-air, and heated again at 220 $\celsius$ for 30 minutes. \\ \\
\endgroup
\begingroup
\textbf{Gas sensing:} For the gas sensing characterization, the samples were mounted inside small chambers, that are connected in series to a gas-flow line, and electrically connected to a digital multimeter (equipped with a multiplexed channel input). Small concentrations of the target gas ($\mathrm{NO_{2}}$) are mixed in the synthetic-air flow, taking the form of step-like pulses with duration of 30-minutes each and separated by a 30-minutes recovery period (Fig. \ref{Fig.resp}a - \textit{gray shaded areas}). Calibrated mass-flow controllers are used to maintain the constant flow of 2-liter/minute in the gas-line and a relative humidity (RH) in the gas mixture during measurement. The set-up (gas characterization bench) is composed by 4x independent gas-lines: (1) $\mathrm{O_{2}}$, (2) $\mathrm{NO_{2}}/\mathrm{N_{2}}$ (2.6 ppm), (3) \textquotesingle dry\textquotesingle-$\mathrm{N_{2}}$ and (4) \textquotesingle wet\textquotesingle-$\mathrm{N_{2}}$. The relative humidity is controlled using a bubbler (filled with DI-water) connected to an independent $\mathrm{N_{2}}$ gas line (\textquotesingle wet\textquotesingle-$\mathrm{N_{2}}$ line). A constant excitation current of $0.5$ $\mu A$ was applied on the samples electrodes during measurements, and the power dissipated per device was typically in the range $1$-$2.5 \mu W$. These operational characteristics corresponds to ultra-low power applications, where a small/compact energy source can be employed (e.g. in smart sensors applications).  \\ \\ 
\endgroup

\section*{Acknowledgement}

%Please use ``The authors thank \ldots'' rather than ``The
%authors would like to thank \ldots''.

%The author thanks Mats Dahlgren for version one of \textsf{achemso},
%and Donald Arseneau for the code taken from \textsf{cite} to move
%citations after punctuation. Many users have provided feedback on the
%class, which is reflected in all of the different demonstrations
%shown in this document.

The authors discloses support for the research of this work from SPW Research [Convention n° 2010243, BEWARE 2020 - Appel 2] and the company VOCSens [n° BCE 0721.610.714]. B.H. (senior research associate) acknowledge support from the F.R.S.-FNRS. The authors also acknowledge financial support from the ARC project DREAMS (21/26.116).

%\backmatter

%\bmhead{Supporting information} 

\section*{Supporting information}

%A listing of the contents of each file supplied as Supporting Information should be included. For instructions on what should be included in the Supporting Information as well as how to prepare this material for publications, refer to the journal's Instructions for Authors.

%The following files are available free of charge.
\begin{itemize}
  \item Supporting Information: Presents electrical resistance versus temperature measurements that allowed to validate the relations given in equation (1). In addition, examples are shown of application of the conversion formula, with the same set-of-parameters used in Fig.\ref{Fig.resp}-b (\textit{gray-curve}), in completely independent resistance-variation data-sets for \ch{NO2} sensing. Finally, to confirm the applicability of our model beyond PbS NCs networks, we present data for ammonia (\ch{NH3}) sensing using a single polypyrrole (PPy) sensors, based on the conversion formula in (\ref{eq:sconv}), derived for single-type interaction site.

 % \item Filename-2: brief description
\end{itemize}

%%%%%%%%%%%%%%%%%%%%%%%%%%%%%%%%%%%%%%%%%%%%%%%%%%%%%%%%%%%%%%%%%%%%%
%% If you are using classical BibTeX rather than biblatex,
%% remove the \printbibliography and uncomment the \bibliograpy one
%%%%%%%%%%%%%%%%%%%%%%%%%%%%%%%%%%%%%%%%%%%%%%%%%%%%%%%%%%%%%%%%%%%%%
%\printbibliography

%\bibliograpy one

\bibliography{Bibliography.bib} 

\newpage

\pagebreak

\end{document}

\maketitle

\section{Conductance model validation}

Considering a conductive layer, where current is limited by transmission through the intergrain potential barriers. We assume each pair of connected NCs (\textit{i, j}) in the percolation network contributing with a resistance $r_{ij}$ (Fig.3a, in the main text), in the form of (\ref{eq:transfer}), where $\phi_{B_{ij}}$ is the height of the intergrain potential barrier, $S$ is the junction area, $A^{*}$ is the Richardson constant, $T$ is the temperature, k is the Boltzmann constant, $q$ is the electron-charge, and $v_{ij}$ is the potential drop across the interface \cite{Grillo2021}.

\begin{equation}\label{eq:transfer}
   r_{ij} = \frac{v_{ij}}{S A^{*} T^{2}} \cdot \left( \frac{e^{\frac{qv_{ij}}{kT}}+1}{e^{\frac{qv_{ij}}{kT}}-1} \right) \cdot \exp{\left(\frac{\phi_{B_{ij}}}{kT} \right)}
\end{equation}

When a bias voltage, $V$, is distributed over a large number of connected particles in the percolation chains (typically several hundreds to thousands in the case of few-nm nanocrystals between electrodes spaced by a few-$\mu$m), in the limit of $qv_{ij}/kT < 1$ one can approximate $r_{ij}$ (in Fig.3a, in the main text), as shown in (\ref{eq:Srija}). Furthermore, to account for combined inhomogeneities in the geometry and interface-composition we can consider a Gaussian broadening of the denominator ($\delta_D$) and of the barrier height ($\delta_B$), and the mean resistance of the interface of connected NCs, $\langle r_{ij} \rangle$, can be written as in (\ref{eq:Srijb}) (i.e. the effect of the presence of inhomogeneities is always to promote $\langle r_{ij} \rangle$ to higher values as $\bigl< {\exp(\delta_B/kT)}/{1 +\delta_D} \bigr> \geq 1$).

%\begin{equation}\label{eq:rindparticles}
%   r_{ij} \simeq \frac{2k}{S_{ij}A^{*}T}\cdot \exp{\left(\frac{\phi_{B_{ij}}}{kT}\right)}
%\end{equation}\\

\begin{subequations}
\begin{align}
r_{ij} \approx \frac{2k}{S_{ij}A^{*}T}\cdot \exp{\left(\frac{\phi_{B_{ij}}}{kT}\right)}\label{eq:Srija}  \\
\langle r_{ij} \rangle = \frac{2k}{SA^{*}T} \exp{\left(\frac{\phi_{B}}{kT} \right)}  \biggl< \frac{\exp{\left(\frac{\delta_B}{kT}\right)}}{1 +\delta_D} \biggr>\label{eq:Srijb}
\end{align}
\label{eq:rij}
\end{subequations}

\begin{figure}[h!]
\centering
\includegraphics[width=0.6\columnwidth]{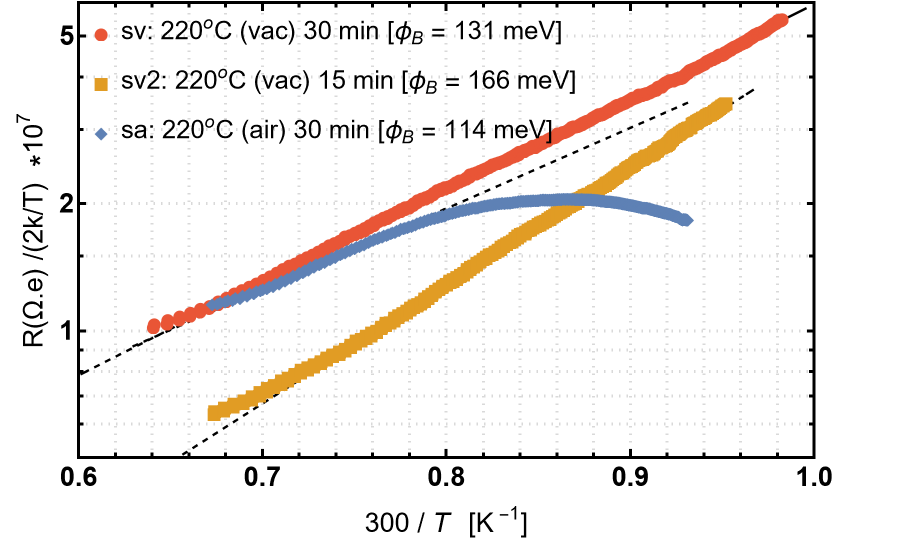}  
\caption{Correspondence between experimental Resistance vs Temperature data for samples \textsf{\textbf{sv}}, \textsf{\textbf{sv2}} and \textsf{\textbf{sa}} and the device model illustrated in Fig.3 (in the main text) where the charge-transport is limited by the intergrain potential barriers formed between PbS-nanocrystals in the sensitive layer. Dashed black lines correspond to fits of the Resistance vs Temperature data using (\ref{eq:Srija}), with the fitted values of potential barriers for each sample indicated in the inset.}\label{Fig.validation}
\end{figure}

To validate equation (\ref{eq:Srija}), i.e. equivalent to equation (1, in the man text), we first checked whether the resistance measured on sensors \textsf{\textbf{sv}}, \textsf{\textbf{sa}}, and on sample \textsf{\textbf{sv2}}, as a function of temperature, obeys the dependence defined in (\ref{eq:Srija}) as assumed in the main text. The additional sample \textsf{\textbf{sv2}} was heat-treated in vacuum, at 220 \textcelsius \ for 15 minutes. The recorded resistances as a function of temperature are presented in Fig.\ref{Fig.validation} (\textit{in red, orange and blue}) together with the respective fitted curves (\textit{black-dashed lines}) using (\ref{eq:Srija}). The experimental curves in Fig.\ref{Fig.validation} were obtained during the cooldown after the thermal treatment, in situ (in the vacuum chamber, for \textsf{\textbf{sv}} and \textsf{\textbf{sv2}}, and in ambient air, for \textsf{\textbf{sa}}). The data points following the model prediction indicate the ranges where the resistances are limited by the potential barriers formed at the interfaces between the grains (as represented in Fig.3a, in the main text). For sample \textsf{\textbf{sa}}, as expected for lower temperatures (in ambient-air), the curve tends to deviate to lower resistance values due to the increase in adsorbed oxygen-gas molecules from the atmosphere, as expected for p-type sensor (i.e., in qualitative agreement with Langmuir's theory of adsorption) \cite{Swenson2019}. We compare the results for the samples measured in vacuum, \textsf{\textbf{sv}} and \textsf{\textbf{sv2}}, indicating that longer annealing times will favor lower potential barriers between grains, probably due to a reduction of the mean distance between the NCs or a relative reduction of Fermi-level pining at the NCs interfaces \cite{Asubar2021}. When comparing between the barrier heights of samples \textsf{\textbf{sv}} and \textsf{\textbf{sa}}, the annealing in ambient-air seems to favor lower barrier heights. The formation of an oxidized interfacial layer can explain the lower potential barrier observed, as the oxygenated species on the surface of p-type semiconductor NCs can form a polar layer that will contribute in lowering the potential barrier \cite{Asubar2021,Choi2014}. The difference in the base-line resistance observed between the samples is attributed to the formation of a nearly random percolation chain between the contact electrodes, and as a result of the semi-controlled deposition method based on drop-drying technique.

\section{Sensing performance}

\begin{figure}[h!]
\centering
\begin{tabular*}{\textwidth}{ll}
{$\mathrm{a}$} & {$\mathrm{c}$} \\
\includegraphics[width=0.5\columnwidth]{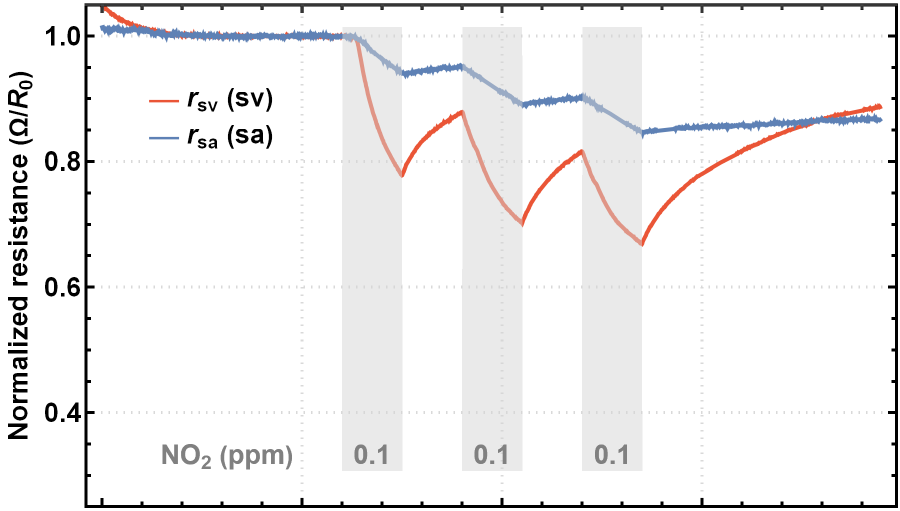} & \includegraphics[width=0.5\columnwidth]{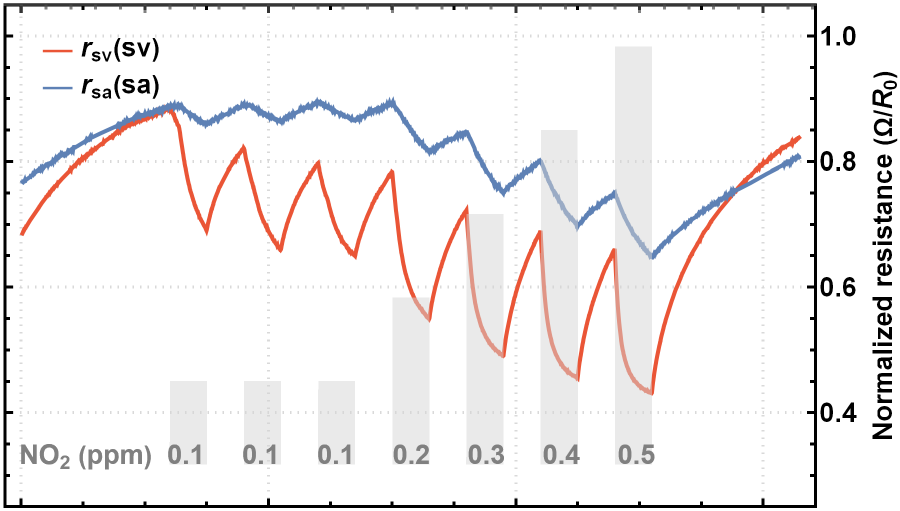} \\
{$\mathrm{b}$} & {$\mathrm{d}$} \\
\includegraphics[width=0.5\columnwidth]{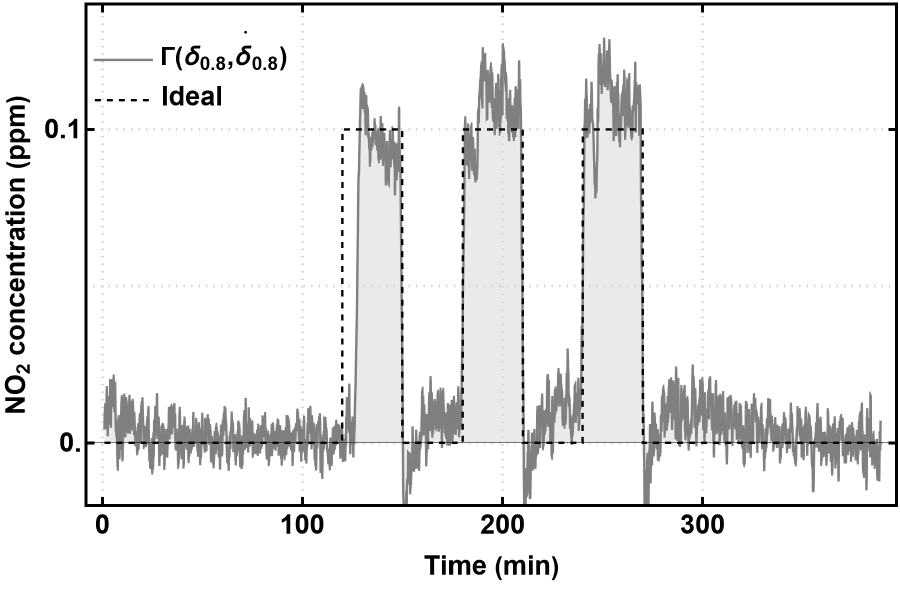} & \includegraphics[width=0.5\columnwidth]{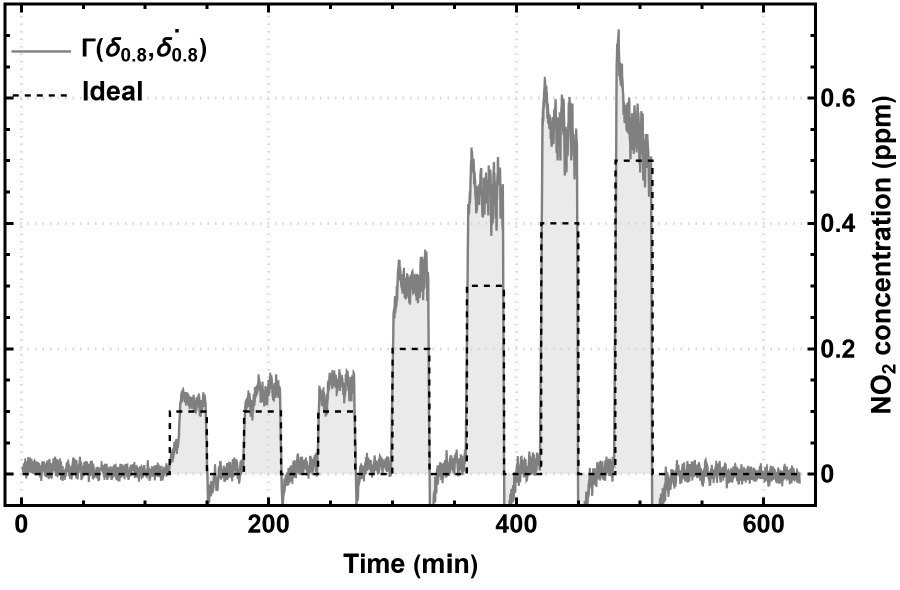} \\
\end{tabular*}
\caption{(a) and (c) Normalized resistance variation measured when different concentrations of \ch{NO_2}-gas are mixed (in intervals of 30-minutes) to a constant flow (2-liters/minute) of synthetic-air with 44\% relative humidity. The \textit{red line} corresponds to sample \textsf{\textbf{sv}} and the \textit{blue line} corresponds to sample \textsf{\textbf{sa}}. In (b) and (d), the (\textit{gray line/shaded area}) shows the converted \ch{NO_2} concentration calculated using the formula in (6, in the main text) with the parameters of Table 2 (main text), and the \textit{black dashed line} corresponding to the ideal (nominal) gas concentration profile applied during the gas-sensing test.}\label{Fig.sens-perf}
\end{figure}

In the Response section (in the main text), the parameters of the conversion formula, $\Gamma(\delta_{b},\dot \delta_{b})$ (6, in the main text), were obtained from fitting of the sensors response in Fig.2a (in the main text), using the device equation (3, in the main text) and the sorption dynamics defined in equation (4, in the main text). This approach corresponds to an indirect calibration yielding the conversion parameters in Table 2 (main text). Thus the conversion formula should allows to convert the electrical resistance measured on both sensors into the actual concentration of target-gas, in real-time. Next, for further evaluation of the sensing performance of the sensor pair, \textsf{\textbf{sv}} and \textsf{\textbf{sa}}, using the double-sensor strategy (DS-strategy), the conversion formula was directly applied on two independent datasets (Figs. \ref{Fig.sens-perf}-a,c). For the first dataset, in Fig.\ref{Fig.sens-perf}-a, the sensors were exposed to 3-identical low concentration step-like pulses (0.1 ppm) of \ch{NO_2}, while in the second dataset, in Fig.\ref{Fig.sens-perf}-c, the sensor-pair was exposed to several step-like pulses between 0.1-to-0.5 ppm, i.e. three initial pulses of 0.1 ppm and linearly increasing to 0.5 ppm. The results for the converted \ch{NO_2} concentration, in Figs \ref{Fig.sens-perf}-b,d, seems to independently validate the application of the conversion formula with the parameters obtained from modeling (Fig.2a, in the Response section in the main text). Overall, a better accuracy and precision seems to be obtained at lower \ch{NO_2} gas concentration, in Fig.\ref{Fig.sens-perf}-b, while for higher concentrations, in Fig.\ref{Fig.sens-perf}-d, one observes a trend to deviate to higher values when compared to the ideal (nominal) concentration in the chamber during the test. Here, the possible presence of a base-line drift due to a ripening effect (e.g. dynamic changes on the surface chemistry) can be considered as the main cause. One can also notice the systematic presence of measurement artifacts, corresponding to slightly negative values reported immediately after suppression of the \ch{NO_2} gas (after each step-like concentration), that can be attributed to the experimental approximation effect in time-derivative $\dot{\delta}(t)$. The precision of the double-sensor strategy applied here is roughly estimated at $20\%$, for the reported gas concentration range, and the noise-floor observed was estimated as $20$-ppb (which can eventually be interpreted as the limit of detection in the frame work of application of the conversion formula (6, in the main text). However, those efficiency figures (accuracy, precision and noise) can certainly be improved, for optimized sensing materials and post-treatments, or depending on the calibration and sensing protocols used: for example, the choice of sensors, as well as the resistance measurement protocol, could be optimized to maximize accuracy and precision, and minimize noise. During the initial stabilization time and during the recovery (after each of the step-like gas-concentration pulses), the resistances in the sensor-pair, $r_{sv}$ and $r_{sa}$, were dynamically varying, in the out-off-equilibrium regime, and it is specially remarkable to note that the conversion formula allowed to compensate the sensors dynamic response and yield \ch{NO_2} concentrations oscillating (closely) near zero before and after each pulse. These results allowed us to confirm the applicability, stability, reproducibility, and accuracy, unlocked by this strategy even for low gas concentrations. 

For completeness, is worth mentioning that the gas sensing tests, in Figs. 2 (in the main text) and \ref{Fig.sens-perf}, were performed approximately 21-days after the samples were manufactured, while in that period they were stored in ambient conditions. After being manufactured, the sensitive layers experiences a ripening period, that can typically be of a few-weeks to a few-months before it can present low baseline drift and stable gas-sensitivity (e.g., on aged samples the recovery time after been exposed to the target gas increases significantly when compared to recently manufactured samples).

%In the supplementary section are presented the gas sensing results, using the sensors $\mathrm{s2}$ and $\mathrm{r1}$, when both were age of more than 1-year, and a comparison is shown between the results obtained with a single-sensor (SS) versus a double-sensor (DS) strategy, using the respective conversion formulas (\ref{eq:7}) and (\ref{eq:8}).  First, the conversion equations were directly applied on a calibration dataset (Figs. S.1 a,b) and the effectiveness of the calibration parameters were tested on a second (independent) dataset (Figs. S.1 c,d). This results (in the supplementary information) validate and confirm the applicability of the dynamic model and the sensing strategy here proposed on aged samples. 

%by using (\ref{eq:8}), and substituting the same set of parameters as in Fig. \ref{fig.4}b, to determine the gas concentration using two independent datasets. The double-sensor strategy was validated by direct application of (\ref{eq:8}) on the the resistance datasets in Figs. \ref{fig.5}a and \ref{fig.5}c. 

% It is remarkable to verify that by applying this sensing strategy allows to compensate the sorption dynamics to enable the measurement of the $\mathrm{NO_{2}}$ gas concentration without delay and at low concentrations (i.e. in the range it poses threats to human health and the environment). 

\section{Ammonia sensing with single-sensor}

%In the case of a single type interaction site on the surface, we can find an analytical formula to directly convert, in real-time, the sensor response, $R(t')$, to a gas concentration, $\Gamma(t')$, i.e., by taking the time-derivative of (\ref{eq:transd}), substituting ${\partial\theta}/{\partial t}$ by (\ref{eq:sorp}) and using $\theta = \left(kT / \alpha \right) \ln{(R_{0}/R)}$ we find the conversion formula in \ref{eq:sconv}. This approach was experimentally validated for ammonia sensing, using a sensor based on a conductive polymer (polypyrrole). In this case the sensing principle involves the Schottky barriers that are formed in the junctions between the doped PPy layer and the gold-electrodes. The set-of-parameters, $A$, $D$, $\alpha$ and $R_{0}$, was obtained from calibration by fitting the conversion formula directly on the response curves (the results are presented in the supplementary).

%\begin{subequations}
%\begin{align}
%r=R/R_{0} \label{eq:sconva}  \\
%\Gamma (r,\dot r) =-\frac{D \ln r+\frac{\dot r}{r}}{A\left[\frac{\alpha}{kT}+\ln r\right]} \label{eq:sconvb}
%\end{align}
%\label{eq:sconv}
%\end{subequations} 

\begin{figure}[h!]
\centering
\begin{tabular*}{\textwidth}{ll}
{$\mathrm{a}$} & {$\mathrm{c}$} \\
\includegraphics[width=0.5\columnwidth]{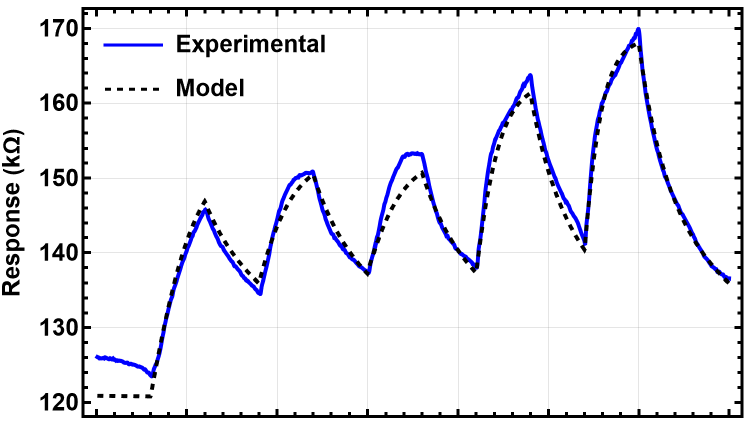} & \includegraphics[width=0.5\columnwidth]{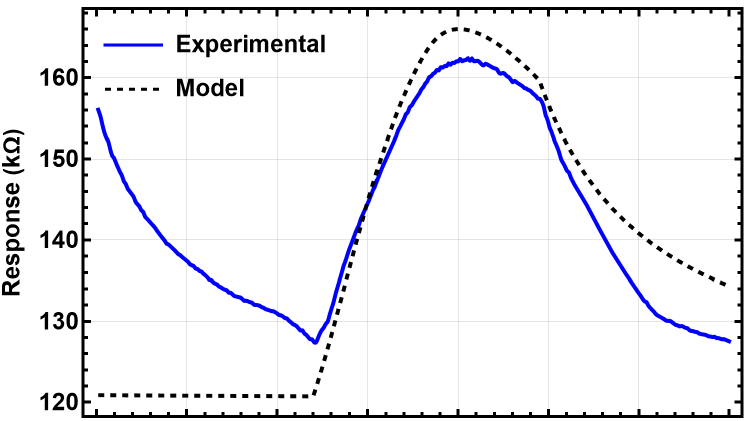} \\
{$\mathrm{b}$} & {$\mathrm{d}$} \\
\includegraphics[width=0.5\columnwidth]{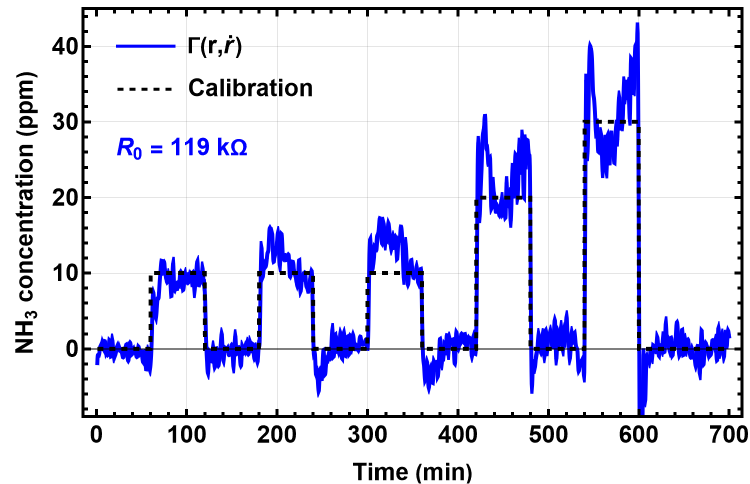} & \includegraphics[width=0.5\columnwidth]{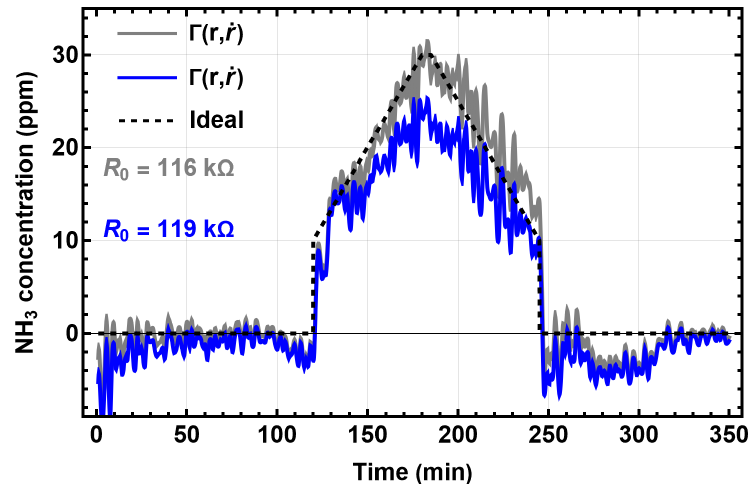} \\
\end{tabular*}
\caption{(a) and (c) Resistance variation measured when different concentrations of ammonia-gas are mixed to a constant flow (2-liters/minute) of synthetic-air with 40-44\% relative humidity, at room-temperature. The \textit{blue lines} corresponding to the experimentally measured values during the tests, where the \textit{black-dashed lines} correspond to calculated values, using the sensor model equations (3, in the main text) and (\ref{eq:amsorp}) and the values in Table \ref{tab.AmModParam}.  In panels (b) and (d), the \textit{blue lines} representing the converted (measured) ammonia concentration profile during the tests, i.e. calculated using (\ref{eq:amsconv}) with the parameters of Table \ref{tab.AmModParam}, and the comparison with the \textit{black dashed line} corresponding to the respective ideal (nominal) gas concentration profile applied during the test. The \textit{gray line} in panel-d shows the better accuracy that is obtained when using a smaller value of the base-line resistance. }\label{Fig.amm}
\end{figure}

In this section, the single-sensor strategy is validated for real-time ammonia sensing, based on an electrodeposited layer of polypyrrole (PPy) on gold electrodes. The sensing mechanism involves the modulation of the Schottky barriers height, that are formed at the junctions between the doped PPy layer and the gold-electrodes \cite{NGUYEN1999}. The sensor was fabricated by electropolymerization, directly on the CMOSEnvi dies (Fig.1a, in the main text), following the synthesis protocol described in Lakner et al \cite{Lakner2020} (the sensor was fabricated on the big IDE, with 500 $\mathrm{\mu m}$ x 1050 $\mathrm{\mu m}$). The electrical resistance of the sensing layer was recorded over time, for two independent datasets, using distinct ammonia concentration profiles, as shown in Figs.\ref{Fig.amm}-(a,c) (blue curves). One of the tests, corresponding to Figs.\ref{Fig.amm}-(a,b), consisted of a sequence of 5x step-like concentrations of ammonia, varying between 10-to-30 ppm, and the other test, in Fig.\ref{Fig.amm}-(c,d), consisted in the introduction of a linearly varying concentration of ammonia, between 10-to-30 ppm. The ammonia gas was introduced in a constant flow of synthetic air at room-temperature, and at constant relative humidity (40-44\%). The respective ammonia concentration profiles applied during these tests are presented in Figs.\ref{Fig.amm}-b,d (black-dashed curves).

In order to validate the single-sensor strategy, the conversion formula (eq.(\ref{eq:amsconv}) below) was first directly calibrated, using the experimental resistance variation in Fig.\ref{Fig.amm}-a (blue curve) as input file, and finding the parameters, $A$, $D$, $\frac{\alpha}{kT}$ and $R_{0}$, that provided the best fitting to the ammonia-concentration profile in Fig.\ref{Fig.amm}-b (black-dashed curve). This calibration protocol provided the values in Table \ref{tab.AmModParam}, where the quality of the fitting can be verified in Fig.\ref{Fig.amm}-b (in blue), and directly compared to the ammonia concentration profile used for calibration (black-dashed curve). As observed in the main text (for the double-sensor strategy), by applying the conversion formula the gas concentration can be extracted quantitatively in real-time, i.e. it is capable of compensating for the slow reaction-recovery, as observed in figs\ref{Fig.amm}-a. However, the conversion formula in (\ref{eq:amsconvb}) differs from that in the main text (5b), for \ch{NO_{2}} sensing based on PbS-NCs, as in the case of ammonia the desorption from PPy is frequently described by a pseudo-second-order kinetics (\ref{eq:amsorp}) which differs from the simple first-order dynamics assumed before (4, in the main text).

The calibration was then verified on the independent dataset, by applying the calibrated conversion formula to the resistance variation (blue curve) in Fig.\ref{Fig.amm}-c. The result for the converted gas concentration can be verified in Fig.\ref{Fig.amm}-d (blue curve). Despite the good match concerning the linear trend of the ammonia concentration profile, a systematic deviation can be observed yielding underestimated ammonia concentration values. This effect was mitigated by replacing the calibrated base-line resistance value, $R_{0} = 119k \Omega$, by a smaller value, i.e. $116k \Omega$ (gray curve, in Fig.\ref{Fig.amm}-d). The drift of the base-line resistance is commonly observed in newly fabricated samples, as an aging effect (with the resistance drifting to higher values), and is consistent with the fact that the measurement in Fig.\ref{Fig.amm}-c (blue curve) was, actually, carried out 4-days before the measurement in Fig.\ref{Fig.amm}-a. In addition, as a further validation, the simulated response using the sensor response-model is plotted in Figs.\ref{Fig.amm}-(a,b) (black-dashed curves), i.e. calculated using equations (3, in the main text) and (\ref{eq:amsorp}), and the parameters in Table \ref{tab.AmModParam} bellow.

%The gas concentration curves, that were obtained after conversion were represented in Figs.\ref{Fig.amm}-a,b,c in blue.

%\begin{subequations}
%\begin{align}
%r=R/R_{0} \label{eq:sconva}  \\
%\Gamma (r,\dot r) =-\frac{D \ln r+\frac{\dot r}{r}}{A\left[\frac{\alpha}{kT}+\ln r\right]} \label{eq:sconvb}
%\end{align}
%\label{eq:sconv}
%\end{subequations} 

\begin{subequations}
\begin{align}
r=R/R_{0} \label{eq:amsconva}  \\
\Gamma (r,\dot r) = \frac{(D/\frac{\alpha}{kT}) \ln^{2} r-\frac{\dot r}{r}}{A\left[\frac{\alpha}{kT}+\ln r\right]} \label{eq:amsconvb}
\end{align}
\label{eq:amsconv}
\end{subequations}

\begin{equation}\label{eq:amsorp}
\frac{\partial\theta}{\partial t} = A \Gamma (1-\theta) - D \theta^{2}
\end{equation}

\begin{table}[h!]
\centering
\caption{Parameters used in (\ref{eq:amsconvb}), to obtain the results in Figs.\ref{Fig.amm}-(b,d) (blue curves).} \label{tab.AmModParam}
\begin{tabular}{lccc}
\hline
$A$ & $D$ & \ \ $\alpha/kT$ & $R_{0} \ (\Omega)$ \\
\hline
0.0011  &  0.024 & -0.52 & 119k \\
\hline
\end{tabular}
\end{table}

%%%%%%%%%%%%%%%%%%%%%%%%%%%%%%%%%%%%%%%%%%%%%%%%%%%%%%%%%%%%%%%%%%%%%
%% If you are using classical BibTeX rather than biblatex,
%% remove the \printbibliography and uncomment the \bibliograpy one
%%%%%%%%%%%%%%%%%%%%%%%%%%%%%%%%%%%%%%%%%%%%%%%%%%%%%%%%%%%%%%%%%%%%%
%\printbibliography

%\bibliograpy one

\bibliography{SuppBibliography} 

\newpage

\pagebreak